\documentclass[transmag,letterpaper]{IEEEtran}
\usepackage{nopageno}
\newif\ifpeerreview

\peerreviewfalse

\newcommand{\paperID}{74}

\usepackage[switch]{lineno}
\usepackage{cite}
\usepackage[hidelinks, bookmarks=false]{hyperref}
\usepackage{amsmath,amssymb,amsfonts}
\usepackage{algorithmic}
\usepackage{graphicx}
\usepackage{textcomp}
\usepackage{xcolor}
\def\BibTeX{{\rm B\kern-.05em{\sc i\kern-.025em b}\kern-.08em
    T\kern-.1667em\lower.7ex\hbox{E}\kern-.125emX}}

\newcommand{\overbar}[1]{\mkern 1.5mu\overline{\mkern-1.5mu#1\mkern-1.5mu}\mkern 1.5mu}
\newcommand{\rect}{\mbox{rect}}
\newcommand{\floor}[1]{\lfloor #1 \rfloor}

\begin{document}

\ifpeerreview
  \linenumbers
  \linenumbersep 5pt\relax
\fi

\title{Video from Stills: \\Lensless Imaging with Rolling Shutter}

\ifpeerreview
\author{Anonymous ICCP 2019 submission \\
Paper ID \paperID}
\else
\author{\IEEEauthorblockN{Nick Antipa$^\dagger$*, Patrick Oare*, Emrah Bostan, Ren Ng, Laura Waller}
\IEEEauthorblockA{\textit{Department of Electrical Engineering and Computer Sciences} \\
\textit{University of California, Berkeley}\\
*Authors contributed equally \\
$\dagger$ nick.antipa@eecs.berkeley.edu}%

}

\fi

\ifpeerreview
\markboth{Anonymous ICCP 2019 submission ID \paperID}%
{}
\else
\fi

\IEEEtitleabstractindextext{
\begin{abstract}
Because image sensor chips have a finite bandwidth with which to read out pixels, recording video typically requires a trade-off between frame rate and pixel count. Compressed sensing techniques can circumvent this trade-off by assuming that the image is compressible. Here, we propose using multiplexing optics to spatially compress the scene, enabling information about the whole scene to be sampled from a row of sensor pixels, which can be read off quickly via a rolling shutter CMOS sensor. Conveniently, such multiplexing can be achieved with a simple lensless, diffuser-based imaging system. Using sparse recovery methods, we are able to recover 140 video frames at over 4,500 frames per second, all from a single captured image with a rolling shutter sensor. Our proof-of-concept system uses easily-fabricated diffusers paired with an off-the-shelf sensor. The resulting prototype enables compressive encoding of high frame rate video into a single rolling shutter exposure, and exceeds the sampling-limited performance of an equivalent global shutter system for sufficiently sparse objects.

\end{abstract}
\ifpeerreview
\else
\fi
}

\maketitle
\thispagestyle{empty}
\IEEEdisplaynontitleabstractindextext

\section{Introduction}
All digital imaging sensors have a finite bit rate for exporting the digital measurement. This limited bit rate restricts the space-time bandwidth of the system, forcing a trade-off between temporal and spatial resolution. Traditionally, increasing the frame rate while maintaining pixel count requires increasing the chip bandwidth, which is expensive. Compressive video approaches seek to break this trade-off by spatio-temporally compressing the video data prior to exporting the bits, effectively encoding more information into the limited bandwidth. While most work in compressive video has focused on redesigning the readout architecture of CMOS chips, we instead propose a compressive video scheme based on optical multiplexing using a diffuser. We demonstrate the concept using a simple lensless camera with an off-the-shelf rolling shutter sensor. Our system effectively encodes 140 frames into a single still image.

\begin{figure}[htbp]
	\centerline{\includegraphics[width=.95 \linewidth]{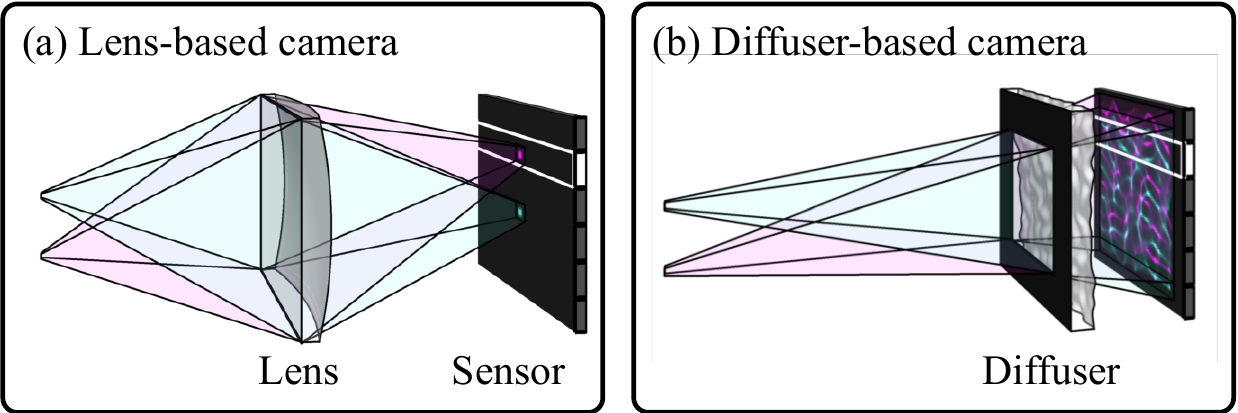}}
	\caption{Diffuser-encoded pseudorandom multiplexing ensures that every row in the sensor measurement contains information from nearly every scene point. (a) A lens-based camera maps each scene point to a point on the sensor. If the sensor samples a subset of rows at a time (outlined in white), as with rolling shutter, only one row of the scene is visible. For example, the cyan point is completely missed in this case. (b) Multiplexing optics, such as a diffuser, spread information across the sensor, allowing the entire scene to be sampled by the subset of rows illustrated here. This effect enables our lensless system to recover a video at a frame rate set by the sensor line scan rate.}
	\label{fig:erasure}
\end{figure}
\vspace{10 mm}

Increasing the frame rate of a sensor with fixed bandwidth can be achieved by reading a subset of pixels at each frame. However, when using one-to-one imaging optics (\textit{i.e.} lenses) that map each scene point to a point on the sensor, information is lost from parts of the sensor that are not sampled. Figure~\ref{fig:erasure}(a) illustrates a sensor with a narrow band of pixels actively recording, placed at the image plane of a lens, with a simple scene consisting of two point sources. The cyan source falls outside of the active exposure band and is therefore not measured. To solve this problem, we propose using spatial-multiplexing optics such that even a small subset of sensor pixels (e.g. one row of a 2D array) contain information from most scene points. Our approach consists of replacing the lens with a pseudorandom phase diffuser placed near the sensor, which maps each point to a distributed, high-contrast pattern of caustics on the sensor. As shown in Fig.~\ref{fig:erasure}(b), the information from every scene point falls on \textit{nearly all} sensor pixels, and is therefore present in the band of rows being read. Recovering a video from a sequence of row measurements then requires solving an underdetermined inverse problem. Because the diffuser produces pseudorandom noise-like measurements, we interpret this as a compressive sensing system, reconstructing the video using sparsity-constrained nonlinear optimization.

To implement this idea, we leverage the ubiquity of \textit{rolling shutter} CMOS sensors. During capture of a single image, rolling shutter sensors expose each row of pixels over a unique time window. This encodes temporal information into the 2D measurement. By randomly multiplexing the scene onto such a sensor, we can recover a video of a dynamic scene wherein each frame corresponds to a row of the rolling shutter capture.

\begin{figure}[htbp]
\centerline{\includegraphics[width=0.97 \linewidth]{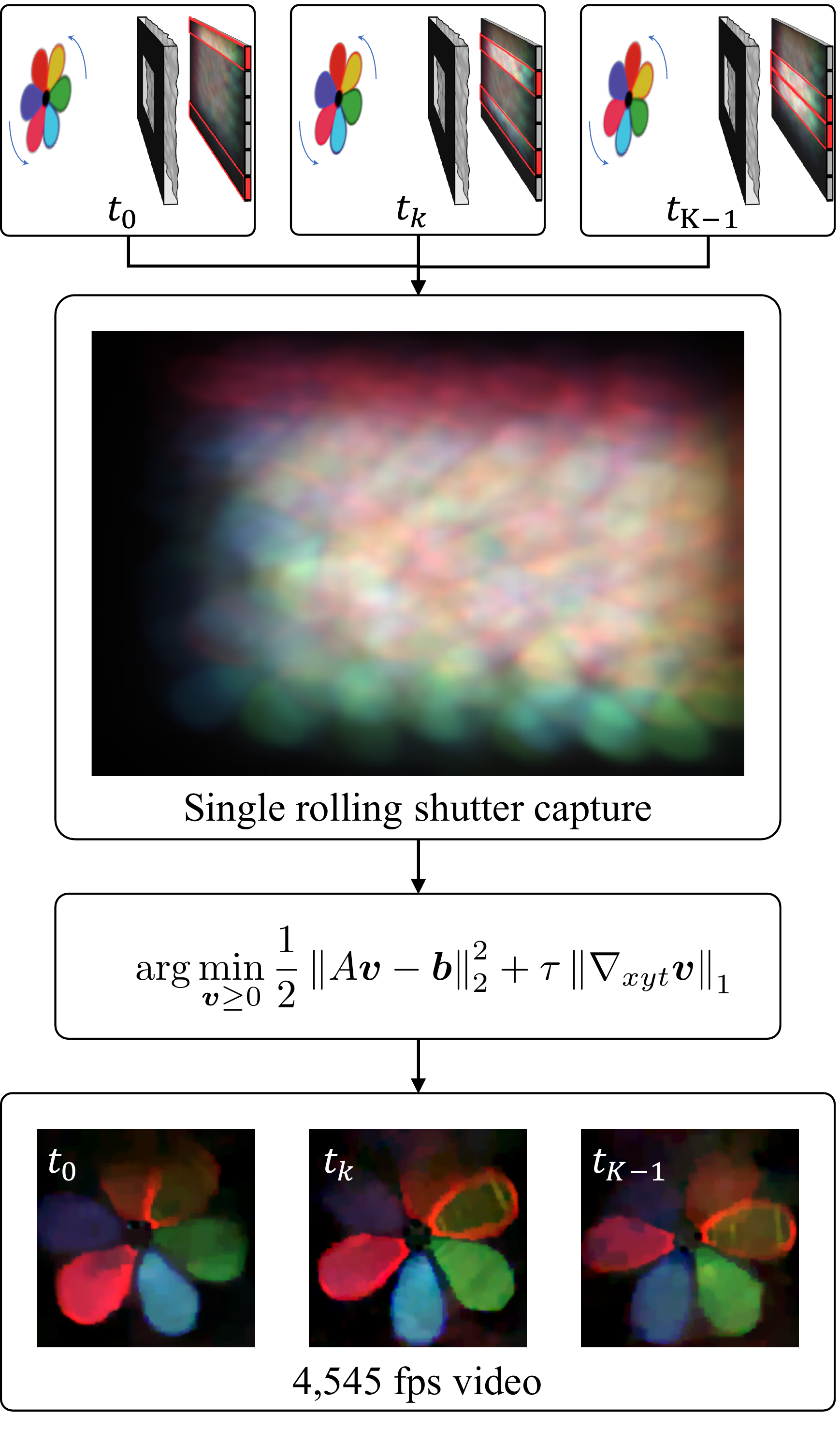}}
\caption{High-speed video from a single-shot rolling shutter image captured by a lensless computational camera. Each row of the recorded image, $\boldsymbol{b}$, is captured at a unique time and contains information about nearly all scene points due to the inherent multiplexing of our lensless imager. The optics and exposure process can be described by a linear forward model, $A$, which is used to solve for the time sequence of 2D images (video), $\boldsymbol{v}$, via non-negative least squares with a 3D gradient sparsity penalty, $\|\nabla_{xyt}\boldsymbol{v}\|_1$, weighted by $\tau$. Each frame of the raw 33 fps recording is expanded to 140 frames giving an effective frame rate of 4,545 fps.}
\label{fig:main}
\end{figure}

Our experimental prototype recovers 140 frames of video at $4,545$ frames-per-second (fps) from a single 2D rolling shutter capture. The system is built using a dual-shutter sCMOS sensor (Fig.~\ref{fig:main}). We analyze the spatial and temporal resolution of the system and show that, for sparse scenes, the spatial resolution significantly surpasses that of much more expensive global shutter approaches at comparable frame rates.

\section{Related work}
To capture high-speed dynamics with conventional sensors, one must overcome the bandwidth limit of digital imaging chips. Compressive video works by spatio-temporally coding the video data prior to capture. Rather than capture a video, then compress it to exploit redundancies, compressive video does the compression step in hardware and captures only relevant data. For example, Hitomi \textit{et al.} proposed a compressive video acquisition scheme that reconstructs a high-speed video from a single image ($9-18\times$ temporal upsampling at 1000 fps)~\cite{Hitomi.etal2011}. The approach relied on pixel-wise programmable exposure timing to modulate the recorded image temporally during the acquisition. Reconstruction was performed through a dictionary of space-time signal patches that is learned offline. Experimentally, the approach used a spatial light modulator (SLM) and global shutter sensor, but could theoretically be implemented on-chip in a CMOS architecture. Using strobed exposure with unique sequences, Veeraraghavan \textit{et al.} reconstructed a high-speed video of periodic events at 2000 fps from a video captured by a camera operating at 25 fps~\cite{Veeraraghavan.etal2011}. Another technique, proposed by Llull and Yuan \textit{et al.}, achieved high-speed video reconstruction (22 frames at 660 fps) from a single-shot coded-aperture image that is obtained by translating binary amplitude masks within the focal plane of a global shutter sensor~\cite{llull2013coded, yuan2014low}. Koller \textit{et al.} later improved the mask design~\cite{Koller.etal15} and Liu \textit{et al.} proposed a reconstruction that exploits the low-rank structure of the underlying scene~\cite{Liu.etal2018}. The commonality between these setups is that each pixel is \textit{temporally} modulated during the exposure, and all require bulky and expensive hardware. Our technique, in contrast, uses simple optics and \textit{spatial} multiplexing rather than temporal.

Rolling shutter can induce undesirable artifacts when imaging dynamic scenes. Removal of such artifacts is an active field of study. Liang \textit{et al.} characterized and corrected the geometric distortions~\cite{Liang.etal2008}. Saurer \textit{et al.} considered extensions for stereo imaging and registration with rolling shutter cameras~\cite{saurer2013rolling}. When camera motion exists, Su and Heidrich~\cite{Su.Heidrich2015} proposed an approach to reconstruct a sharp image by simultaneously removing the motion blur and rolling shutter distortions. 

Rather than undoing the effects of rolling shutter sensors, we seek to leverage them for performance. Gu \textit{et al.} have proposed controlling the readout timing and exposure length for each row~\cite{Gu.etal2010} such that the exposure time discrepancy in subsequent rows enables one to flexibly sample the 3D space-time volume of the dynamic scene. In simulations, their architecture-level proposal was beneficial for computational photography applications such as high dynamic range (HDR) imaging and auto-exposure, but did not successfully resolve video using sparse recovery methods. Oieke and Gamal proposed another architecture that used spatial multiplexing at the chip-level, which allowed them to reach 1920 fps data rate for 256$\times$256 pixel count. Another method uses digital micro-mirror devices (DMDs) for aperture coding and streak cameras with femtosecond speeds to reconstruct ultrafast videos (10 trillion fps) from a single image~\cite{Gao.etal2014, Liang.etal2018}. Liu \textit{et al.} considered similar ideas and used a galvanometer to perform streaking (\textit{i.e.} temporal shearing of the scene)~\cite{liu2019spatial_coded_galvo}. While this concept is similar to ours in spirit, they do not consider spatial multiplexing and they rely on complex, costly hardware. Finally, Sheinin \textit{et al.} recently used rolling shutter and spatial multiplexing to detect and de-mix the contributions from flickering light bulbs in a scene, providing useful information about the power grid. The authors observed that spatial-multiplexing via a diffuser enabled observation of spatio-temporal information, but they do not considering high-speed imaging directly~\cite{iccp2018_Sheinin_rollingshutter_grid}. 

Spatially-multiplexed image capture has been a key ingredient for compressive imaging~\cite{Duarte.etal2008}. Using amplitude masks, Salman \textit{et al.} realized such ideas on a lensless and compact system~\cite{asif2017flatcam}. Diffuser (\textit{i.e.} phase mask)-based lensless cameras have been shown to be capable of 2D imaging \cite{kuo20172D_diffusercam}, and single-shot 3D imaging \cite{antipa2018diffusercam}. Here, we show that diffusers are useful optical elements for compressive video systems, allowing each frame of video to be sampled from a small subset of sensor pixels. Our system can be calibrated from a single image, fabricated using simple lab equipment, and reconstructed using computationally-efficient convolution-based algorithms. 

\section{Forward model and inverse problem}

In this section, we outline a forward model for the optics and the rolling shutter exposure, as well as the inverse problem approach. We will use this model to analyze the temporal resolution of the system in Section~\ref{sec:res_theory}.

\subsection{Rolling shutter model}
\begin{figure}[tbhp] 
\centerline{\includegraphics[width=0.97 \linewidth]{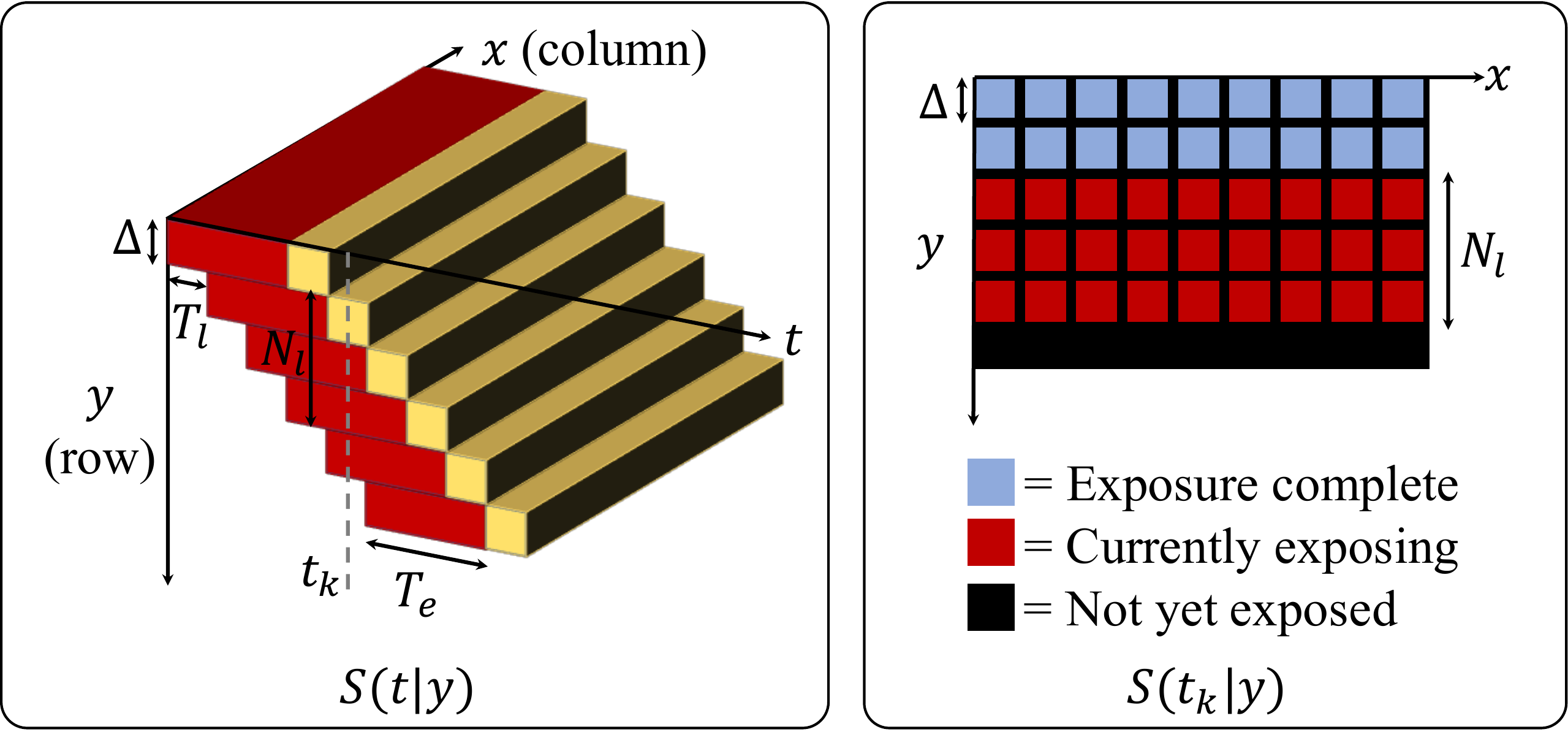}}
\caption{(Left) Spatio-temporal illustration of the rolling shutter function $S(t|y)$ for a sensor with pixel size $\Delta$ and exposure time $T_e$. Red depicts active exposure, and gold is the readout time. (Right) A slice through $S$ at time $t_k$. Each row begins exposing $T_{l}$ seconds after the previous row begins, with red representing actively exposing rows, and blue representing completed rows. The number of rows simultaneously exposed is $N_l=T_e/T_l$, which in this example is 3. For simplicity, we choose $T_e$ such that $N_l$ is an integer.}
\label{fig:timing}
\end{figure}

In general, the exposure at each point on the sensor, $L(x,y)$, can be modeled as a temporal integral, 

\begin{equation}\label{eq:rolling_model}
L(x,y) = \int_{0}^\infty S(t|x,y) \cdot \tilde{v}(x,y,t) dt,
\end{equation}

\noindent where $\tilde{v}(x,y,t)$ represents the time-varying optical intensity on the sensor, and $S(t|x,y) \in \{0,1\}$ is a 3D indicator, the \textit{shutter function}, that encodes the temporal exposure window at each $(x,y)$ position. While our approach could be generalized to different exposure patterns, we focus on rolling shutter due to its ubiquity. Rolling shutter is a column-parallel approach in which each row of pixels exposes for $T_e$ seconds, beginning at a delay, $T_l$, after the previous row began (typically tens-of-microseconds). Because rolling shutter records row-by-row, we drop the $x$-dependence of the shutter function, denoting it as $S(t|y)$ for the remainder of the paper. At any given instant, a small band of $N_l=T_e/T_l$ rows is actively recording photons. For a sensor with pixel size $\Delta$, this is depicted in Fig.~\ref{fig:timing}, with red indicating where $S(t|y)=1$. Our goal is to spatially multiplex scene information into the exposure band at each time point, which enables each band to produce a frame of the final video, achieving frame rates equal to $1/T_l$ fps. 

\subsection{Lensless imaging model}
\begin{figure}[tbhp]
	\centerline{\includegraphics[width=1 \linewidth]{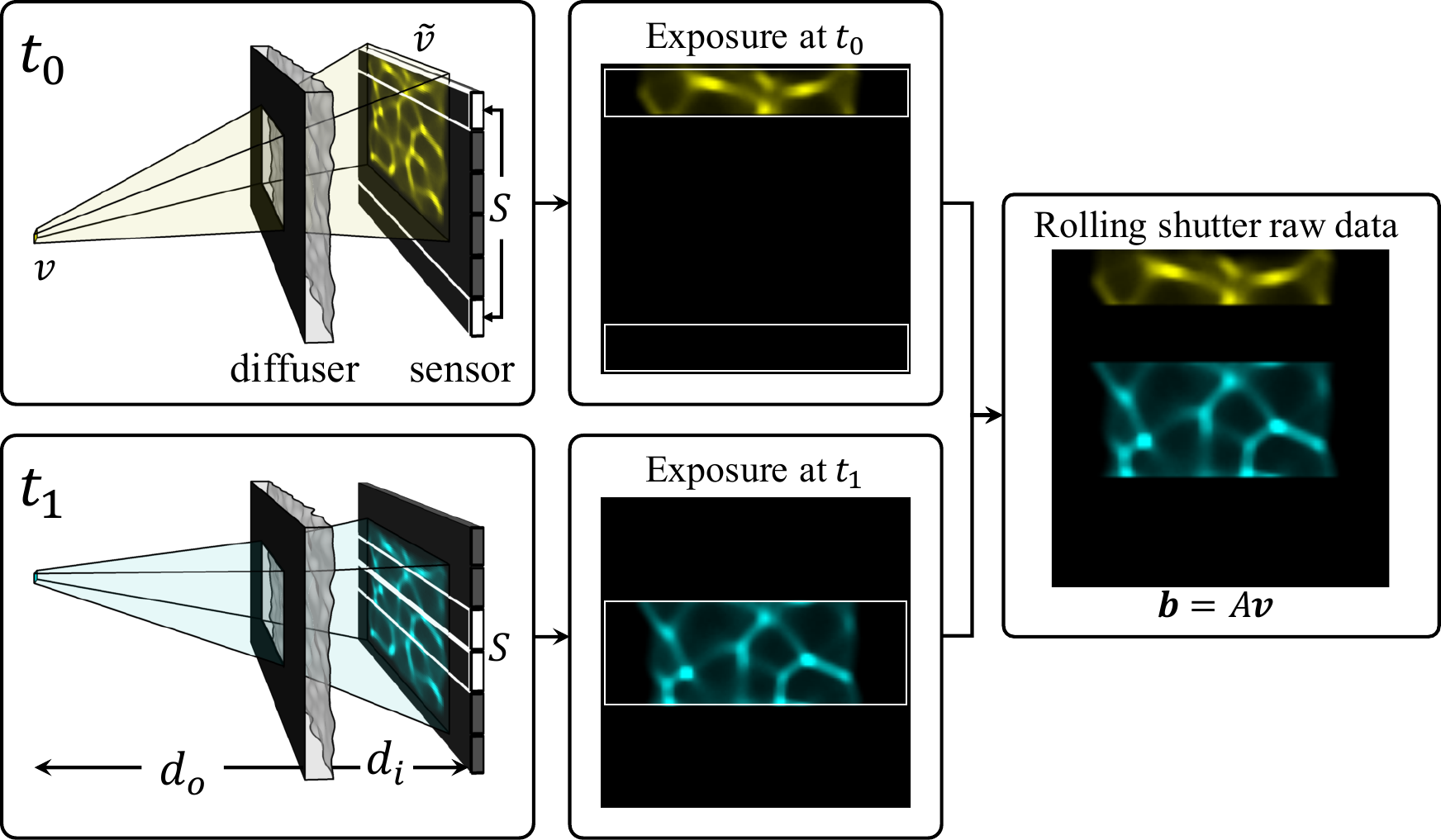}}
	\caption{Image formation for a time-varying scene with two point sources (one yellow, one blue) flashing at unique $y$ locations and times $t_0$ and $t_1$. (Left) Data measurement at times $t_0$ and $t_1$, with the time varying optical intensity, $\tilde{v}(x,y,t_i)$ rendered on the sensor, and dual shutter function $S(t_i|y)$ outlined in white. (Middle) The instantaneous exposure $S(t_i|y)\cdot \tilde{v}(x,y,t_i)$, is shown for each point source. (Right) The captured rolling shutter image is their sum. Due to the spatially-multiplexed optics, nearly all scene points project information into $S(y,t)$. This provides enough information to recover a video from a single image by solving an inverse problem.}
	\label{fig:forward_model}
\end{figure}

In order to achieve the desired multiplexing, we use a simple lensless architecture (see Fig.~\ref{fig:forward_model}) that employs a diffuser -- a pseudorandom phase optic -- as a computational imaging element~\cite{antipa2018diffusercam, antipa2016_diffuser_light_fields}. The system comprises a diffuser placed a distance $d_0$ from the rolling shutter sensor, with the scene at distance $d_i$ from the diffuser. An aperture placed on the diffuser ensures that the resulting Point Spread Function (PSF) is shift-invariant, and enables simple calibration~\cite{antipa2018diffusercam, antipa2016_diffuser_light_fields}. For magnification $m=d_i/d_0$, the sensor plane intensity can be modeled by convolving the magnified scene intensity, $v({x/m},{y/m},t)$, with $h(x,y)$, the on-axis PSF ~\cite{goodman2005introduction}:

\begin{align}\label{eq:conv_model}
    \tilde{v}(x,y,t)=v\left(\frac{x}{m},\frac{y}{m},t\right)\stackrel{(x,y)}{*}h(x,y),
\end{align}

\noindent where $\stackrel{(x,y)}{*}$ denotes linear convolution over $(x,y)$. The diffuser's PSF fills nearly the entire sensor with a pseudorandom caustic intensity pattern that is unique for each shift. This high degree of spatial multiplexing is key to how our system works, enabling any horizontal slice of $\tilde{v}(x,y,t)$ to contain information about nearly all $(x,y)$ positions in the scene. 

\begin{figure}[tbhp] 
	\centerline{\includegraphics[width=1 \linewidth]{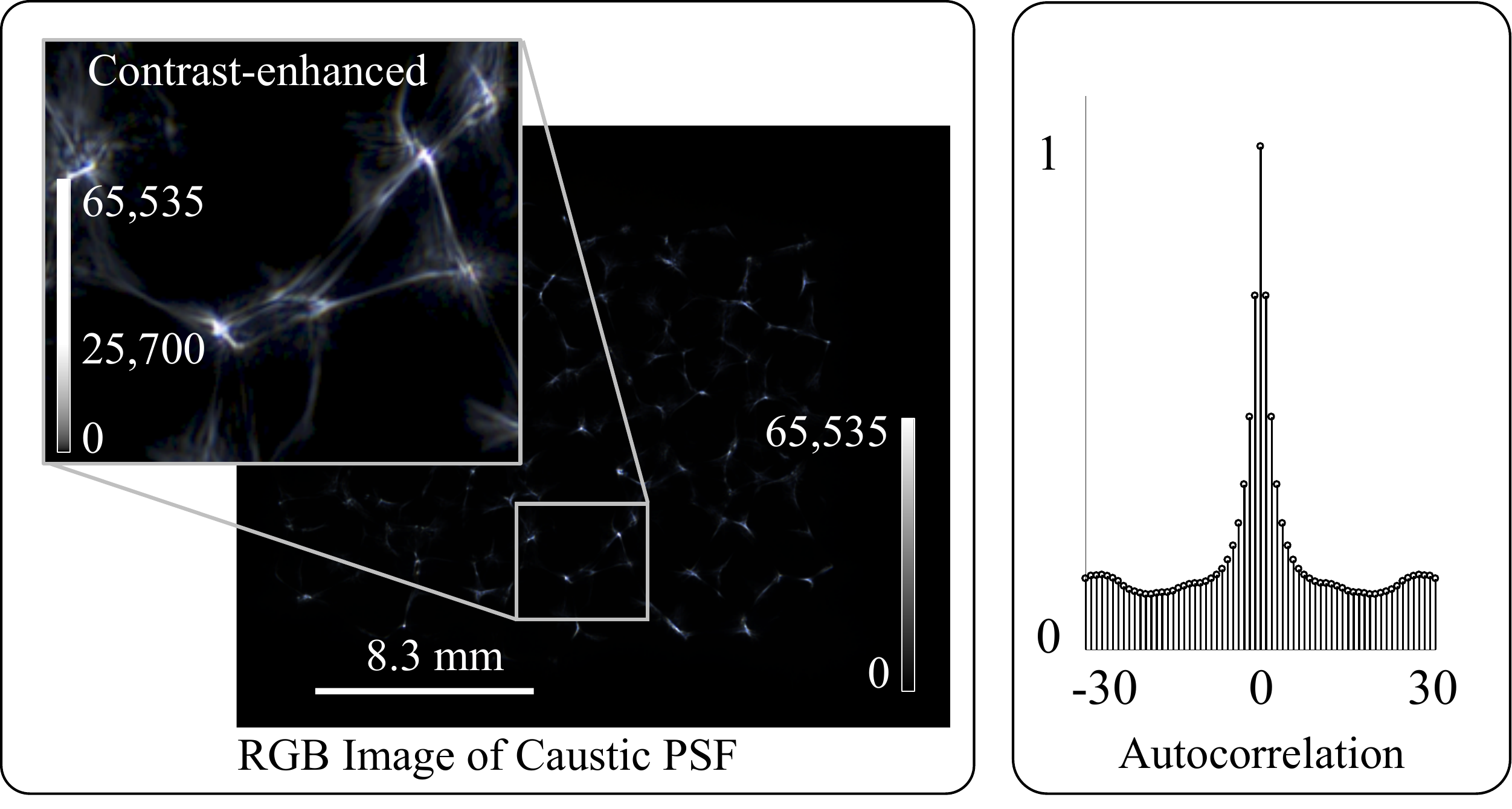}}
	\caption{Left: 16-bit RGB image of the diffuser's caustic point spread function (PSF) for a white LED point source a distance 830 mm from the diffuser. A contrast stretched crop ($\gamma$ = 0.5) is shown inset to show the structure of the caustics. Right: A slice from the normalized autocorrelation of the green channel showing a sharp main peak and relatively low side lobes, making this pattern suitable for compressed sensing.}
	\label{fig:caustics}
\end{figure}

\subsection{Combining lensless and rolling shutter models}\label{sec:combining_lensless_and_rolling}
To solve for the video, we need a discrete forward model. We treat the measurement as a vector of samples taken from the continuous exposure $L(x,y)$: $\boldsymbol{b}[i,j] = L(j \Delta, i \Delta)$, where $i$ and $j$ index the sensor rows and columns, respectively. This leads to a discretized (magnified) scene, denoted $\boldsymbol{v}$, on a 3D spatio-temporal grid with lateral spacing $\Delta$. The temporal spacing is $T_l$, as discussed in section~\ref{sec:res_theory}. This leads to the linear discrete forward model:
\begin{align}
    \boldsymbol{b} &= \sum_{k=0}^{K-1} \overbar{S}_k[i]\cdot \left(h[i,j] \stackrel{[i,j]}{*}\boldsymbol{v}[i,j,k]\right)\\
    &= A\boldsymbol{v}
\end{align}

\noindent where $\stackrel{[i,j]}{*}$ represents discrete linear 2D convolution over the spatial dimensions, $\overbar{S}_k[i] = S(kT_l|i\Delta)$ is the discrete shutter function, and $K$ is the number of recovered frames. Note that for global shutter, this would be a cropped convolution identical to \cite{antipa2018diffusercam,kuo20172D_diffusercam}, but here we absorb the crop into the definition of $\overbar{S}_k[i]$. This linear forward model, denoted $A$ in matrix form, is depicted in Fig.~\ref{fig:forward_model}.

\subsection{Video Recovery}
To recover a video from a single rolling shutter measurement, we must solve an underdetermined linear inverse problem. For a dual-shutter camera such as ours, each symmetric pair of rows in the measurement corresponds to a frame in the reconstruction, so we recover approximately $K = M/2$ frames from a single $M \times N$ capture. The diffuser produces pseudorandom noise-like measurements, so our system fits within the framework of compressed sensing (as demonstrated in \cite{antipa2018diffusercam}). Hence we can solve the underdetermined problem for sparsely-represented scenes using $\ell1$ minimization. We impose a weighted 3D total variation (3DTV) prior on the scene, so the reconstructed video, $\boldsymbol{v}^*$, can be written as the solution to:

\begin{equation}\label{eq:inverse_problem}
\boldsymbol{v}^*=\arg \min_{\boldsymbol{v}\geq 0} \frac{1}{2} \left\| A\boldsymbol{v} - \boldsymbol{b} \right\|_2^2 + \tau \left\| \nabla_{xyt} \boldsymbol{v}\right\|_1,
\end{equation}

\noindent where $\nabla_{xyt} = \begin{bmatrix}\nabla_x & \nabla_y & \alpha \nabla_t \end{bmatrix}^\intercal$ is the matrix of forward finite differences in the $x$, $y$, and $t$ directions. We include an additional tuning parameter, $\alpha$, that weights the temporal gradient sparsity penalty relative to the spatial dimensions (typically set between 5 and 30). We use FISTA~\cite{beck2009_FISTA} with the weighted anisotropic 3DTV proximal operator, implemented using parallel proximal methods according to~\cite{kamilov2017parallel_haar_TV}. For computational efficiency, we never instantiate the matrix $A$ explicitly, but instead compute the matrix-vector products $A(\cdot)$ and $A^H(\cdot)$ using a combination of zero-padding, FFT-based convolutions, and cropping. Each color channel of the video is processed separately, using the corresponding color from the calibrated PSF. This inherently compensates for much of the chromatic aberration in the system.

\section{Experiments}
\subsection{System Design}\label{sec:system_design}

We built our prototype around a PCO Edge 5.5 sCMOS sensor, set to slow-scan rolling shutter mode. The dual shutter reads simultaneously from the top and bottom of the sensor. 

Our homemade diffuser consists of randomly spaced lenslets. Because the lenslets concentrate light into sharp points, random lenslets have been shown to perform well in low-light situations \cite{kuo2018_DiffuserCam_Fluorescence}, as is typical with high-speed imaging. Additionally, the uniformly random lateral placement of the lenslets ensures that each scene point produces a unique pattern on the sensor, and contributes a similar amount of light to each exposure band. This is not true near the edge of the sensor, as discussed in Section~\ref{sec:sampling_and_aperture}.

We fabricate our random lenslet diffusers using the molding process outlined in Section~\ref{sec:prototype_details}. Each lenslet comprising the diffuser has a focal length of $12.7$ mm, yielding an approximately $30^{\circ}$ by $40^\circ$ (width-by-height) half field-of-view (FoV), which is reasonable for photographic scenes \cite{kuo20172D_diffusercam}. The system is calibrated using a single image of a white point source placed in the scene. Figure~\ref{fig:caustics} shows a 16-bit color image of the PSF along with its 2D autocorrelation.

\subsection{Experimental results}
\begin{figure*}[htbp]
	\centerline{\includegraphics[width=1 \linewidth]{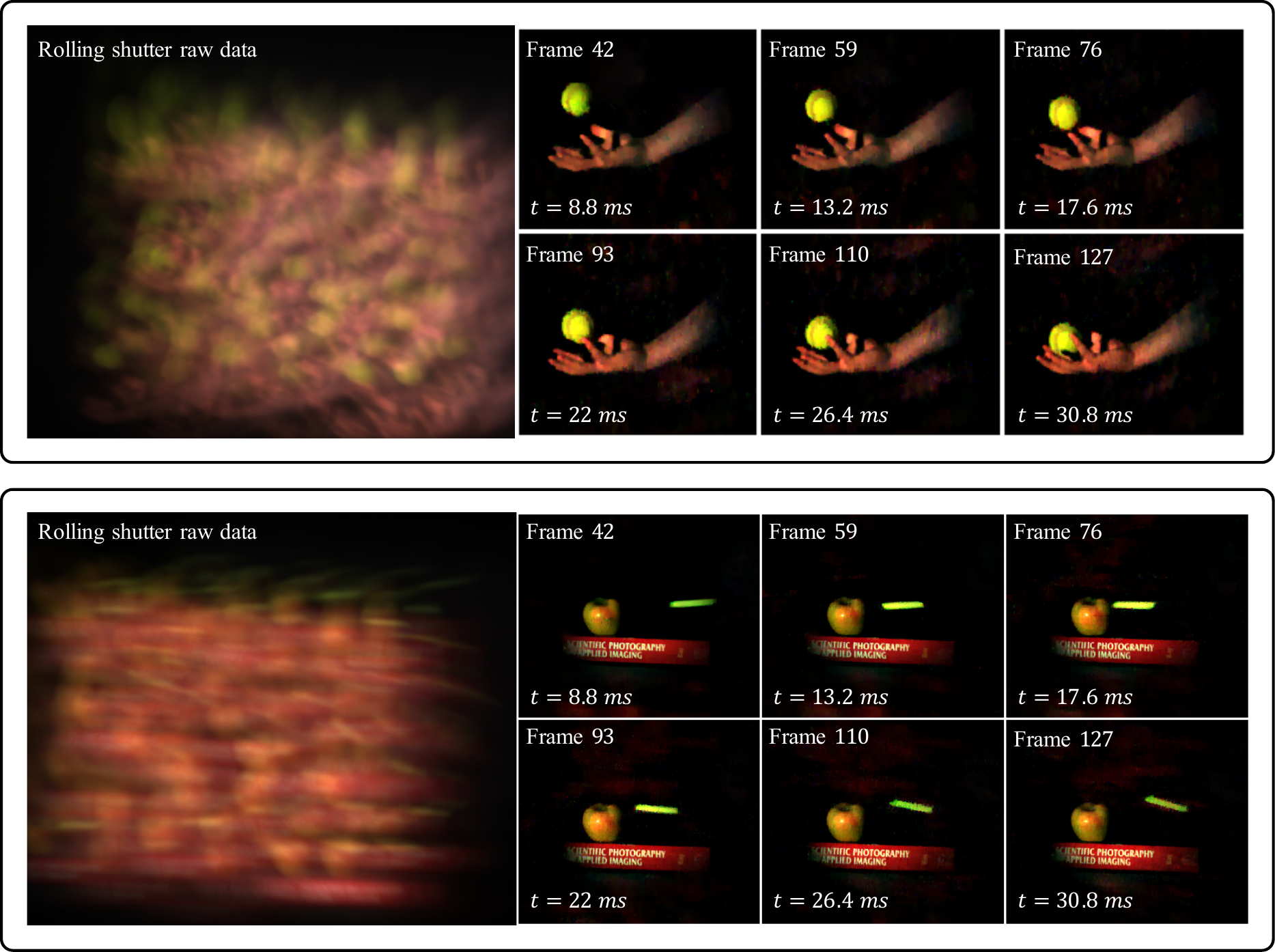}}
	\caption{Experimental videos reconstructed from single-shot images (with 660 $\mu $s exposure). The top example shows a tennis ball falling into a hand, reconstructed with with $8x$ downsampling, and cropped to the center 135 $\times$ 160 pixels (see Supplementary Video 1~\cite{antipa_2019_ICCP_video_supplement}). The bottom example shows a green foam dart ricocheting off an apple with $4\times$ downsampling, cropped to 270 $\times$ 320 (see Supplementary Video 2~\cite{antipa_2019_ICCP_video_supplement}). In both, the raw captured data is shown on the left, with a few frames from the reconstructed video shown at right. The final result contains 140 frames. }
	\label{fig:results_main}
\end{figure*}

To test our system, we captured a variety of dynamic scenes. The raw data is downsampled by either 4$\times$ or 8$\times$ to match the expected temporal bandwidth (see Section~\ref{sec:temp_analysis}). Videos are reconstructed at $640 \times 540 \times 140$ voxel grid for $4\times$ downsampling, or $320\times 270\times 140$ for $8\times$. In both cases the video spans $30.8$ milliseconds. Two example reconstructions are shown in Fig.~\ref{fig:results_main}. The first is a tennis ball dropping into a hand. The second is a green foam dart ricocheting off of an apple placed on a text book. In both cases, motion is clearly visible with good temporal detail present (see Supplementary Videos~\cite{antipa_2019_ICCP_video_supplement}). Due to system geometry, the outer sensor rows are relatively insensitive to the center of the object, degrading the quality of the first 30-40 frames. This is not a fundamental limit of our approach, but is rather a consequence of our implementation (see Sec.~\ref{sec:sampling_and_aperture} for more discussion).

\section{Analysis and Discussion} \label{sec:res_theory}
In this section, we analyze the temporal behavior of the system, showing that the temporal frequencies are band-limited by the exposure time. This motivates the design choices of our prototype, including the diffuser, exposure time, and use of binning (downsampling).

\subsection{Temporal resolution}\label{sec:temp_analysis}

Next, we analyze the temporal frequency content of the measurements to validate temporal resolution. Intuitively, short exposure times are required to achieve high temporal resolution. We will show that, because our system is only compressive in space, its temporal resolution is Nyquist limited, with an inherent band-limit set by the exposure time $T_e$, and the sampling rate determined by the line time, $T_l$. To show this, we begin by writing an expression for $S(t|y)$. As depicted in Fig.~\ref{fig:timing}, $S(t|y)$ is a 1D temporal rectangular window of width $T_e$ seconds, offset by $T_l$ seconds per row:

\begin{equation}\label{eq:S(t;y)}
S(t|y) = \rect\left[\frac{t-\frac{T_{e}}{2} - \floor{y/\Delta} T_l}{T_{e}}\right],
\end{equation}

\noindent where $\floor{y/\Delta}$ represents the row index. Substituting this into the continuous model for rolling shutter acquisition, Eq.~\ref{eq:rolling_model}: 

\begin{equation}\label{eq:continuous_forward}
    L(x,y) = \int \limits_{-\infty}^{\hspace{12 pt}\infty} \rect\left[\frac{t-t_c(y)}{T_{e}}\right] \tilde{v}(x,y,t) dt,
\end{equation}

\noindent where we define $t_c(y) := \frac{T_{e}}{2} + \floor{y/\Delta} T_l$ for compactness. Upon inspection, we see that this is a 1D convolution in the time dimension between the time-varying intensity at the sensor, $\tilde{v}(x,y,t)$, and a rectangular window of width $T_e$. The result of the convolution is evaluated along the slice of 3D space-time defined by $(x,y,t)=(x,y,t_c(y))$: 


\begin{equation}\label{eq:temporal_lowpass}
L(x,y)=\left[\tilde{v} \stackrel{t}{*} \rect\left(\frac{t}{T_e}\right)\right]\bigg\rvert_{\left(x,y,t_c(y)\right)}.
\end{equation}
This captures both the temporal band-limiting inherent in the exposure process as well as the mapping from time to row. Next we substitute Eq.~\ref{eq:conv_model}, the expression for the spatially-multiplexed video, into Eq. \ref{eq:temporal_lowpass}:


\begin{align}
L(x,y)&=\left\{\left[h\stackrel{(x,y)}{*}v_g\right]\stackrel{t}{*}\rect\left(\frac{t}{T_e}\right) \right\}\bigg\rvert_{\left(x,y,t_c(y)\right)}\nonumber\\
&=\left\{h\stackrel{(x,y)}{*}\left[v_g\stackrel{t}{*}\rect\left(\frac{t}{T_e}\right) \right]\right\}\bigg\rvert_{\left(x,y,t_c(y)\right)},\label{eq:full_conv}
\end{align} 

\noindent where $v_g = v(x/m,y/m,t)$ and the convolutions have been reordered, associating the temporal low-pass filter with the input signal. This shows that, while we are multiplexing in space, the temporal information in the system is band-limited by the pixel exposure time.

\begin{figure*}[htbp] 
\centerline{\includegraphics[width=1 \linewidth]{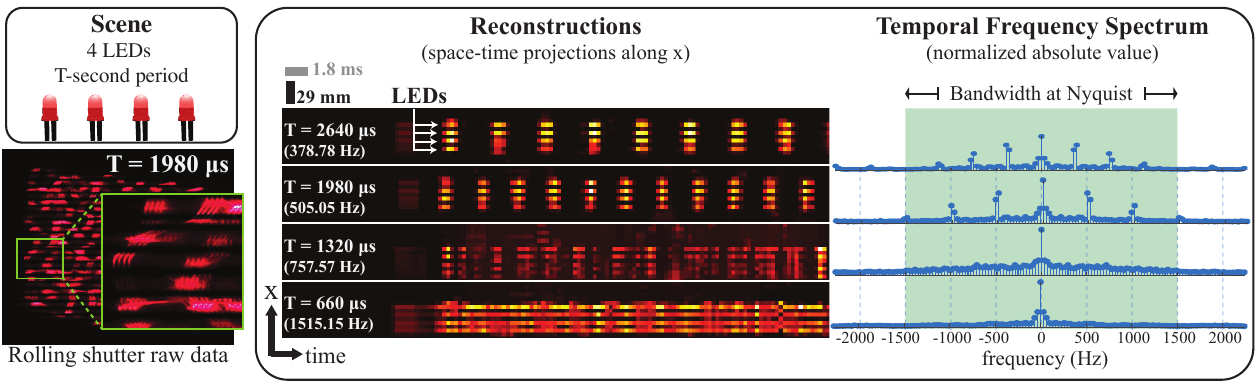}}
\caption{Resolution analysis using a sample consisting of a linear array of 4 LEDs, pulsed synchronously. We vary the pulse frequency of all four simultaneously. (Left) The raw data (660 $\mu s$ exposure time) contains 4 copies of the caustic PSF pattern, each shifted in the horizontal direction according to each LED's spatial position, and the temporal patterns modulate the caustics in the $y$-direction. (Middle) $x-t$ projections of the reconstructed video. As expected, the performance degrades for the LEDs with shorter pulse periods, up to the theoretical limit of 660  $\mu s$ predicted by Eq.~\ref{eq:temporal_lowpass}. (Right) Temporal power spectra of the projections, clearly showing peaks in the time-direction moving as the LED frequency varies.}
\label{fig:res}
\end{figure*}

Finally, we introduce sampling. As shown in Section \ref{sec:combining_lensless_and_rolling}, the measured image is generated by sampling $L(x,y)$ on a grid of spacing $\Delta$. Applying this sampling to the arguments of Eq.~\ref{eq:full_conv}, we get $t_c(y = i \Delta)  = T_l\floor{i\Delta/\Delta} + T_e/2 =T_l i + T_e/2$.  In other words, due to the implicit mapping of time to space, the rolling shutter effectively samples at a rate of $1/T_l$ Hz. Hence we expect to avoid temporal aliasing when  $T_e> 2T_l$, even if the scene contains faster dynamics. This is also why, as discussed in Section \ref{sec:combining_lensless_and_rolling}, we discretize the video on a temporal grid of spacing $T_l$. 

For our sensor, the minimum exposure time is $500$ $\mu s$, with a maximum line time of $27.5 \mu s$. This would result in significant temporal oversampling, which is computationally wasteful. Thus, in practice, we use a combination of lateral downsampling of the raw data and temporal binning of the reconstruction to maintain inter-frame times of $220 \mu s$ (4,545 fps), which better matches the minimum exposure time. Hence we expect to observe dynamics up to $2$ kHz at best. Note that our reconstruction is highly nonlinear, relying heavily on nonnegativity and 3DTV denoising. As a result, this analysis represents only an upper bound to the frequencies we can hope to recover. In practice, measurement noise, calibration error, and regularization reduce performance (see Fig.~\ref{fig:res}). 


\subsection{Resolution validation}
As experimental validation of spatial and temporal resolution, we use a linear array of 4 LEDs flashing in unison with variable frequency square waves. We space the LEDs at the minimum separation resolvable by our system, which we establish empirically by varying the spacing until the LEDs are barely resolved in the reconstructions (6 mm separation at a distance $830$ mm from the diffuser, or $0.4^\circ$ angular resolution). We use an exposure time of $T_e=660 \mu s$, so $N_l=3$ rows are exposing in each band. This should result in maximum frequency of $1,515$ Hz. 

This dynamic scene can be expressed as $v(x,y,t) = u(x,y) \cdot f(t)$, where $u(x,y)$ represents the 2D distribution of LEDs, and $f(t)$ is the modulating waveform. For such an object, the intensity inside the camera body will be $\tilde{v}(x,y,t) = f(t) \cdot \left(h(x,y)*u(x/m,y/m)\right)$.  Plugging this into Eq.~\ref{eq:temporal_lowpass}, we see that the continuous exposure at the sensor will be

\begin{equation}\label{eq:modulation}
	L(x,y) = \left(h\stackrel{(x,y)}{*}u_g\right)\cdot \left[f(t) \stackrel{t}{*} \rect\left(\frac{t}{T_e}\right)\right]\bigg\rvert_{t=t_c(y)},
\end{equation}

\noindent where $u_g = u(x/m,y/m)$. Therefore we expect the measurement to look like the 2D scene convolved with the PSF and modulated in the $y$-direction by the low-pass filtered waveform. Figure~\ref{fig:res} shows raw data from our experimental system. Because the 2D scene is 4 point sources in a line, this appears as 4 laterally shifted copies of the PSF, periodically modulated in the $y$-direction, as expected.


While our analysis provides a bound, experimental errors and nonlinear reconstruction can further deteriorate performance. To test how close we get to the limit, we recorded measurements with LED pulse rates varied from $2,640 \mu s$ (378.78 Hz) to $660 \mu s$ (1,515.15 Hz), the highest frequency predicted by the theory. The results are shown in Fig.~\ref{fig:res}. On the left is a raw measurement with temporal period $T=1,980$ $\mu s$ (505 Hz). A strong envelope is clearly visible, modulating the measurement with a period of $T/T_l = 9$ pixels in the $y$-direction. In the reconstructions, we can clearly resolve all 4 LEDs spatially in all cases. At lower frequencies, the pulses are well resolved in time, with the harmonic structure of the square waves visible in the power spectra. As the period decreases, the temporal contrast reduces, with $660$ $\mu s$ period being totally unresolved.

For comparison, to record the same dynamic scene with LEDs pulsing at $T=1980 \mu s$ using global shutter would require 30 frames at greater than $1,010$ fps. Within our system's sample budget of $270 \times 320=86,400$ samples, each frame from the corresponding global shutter system would only contain $49 \times 58$ pixels. This is a $6\times$ degradation in lateral resolution compared to what our compressive scheme achieves experimentally. Hence, at least for sparse scenes, the compressive approach surpasses a direct sampling scheme. 



\subsection{Diffuser fabrication}\label{sec:prototype_details}

Based on simulations, we found that a diffuser consisting of randomly-spaced lenslets performed better than off-the-shelf diffusers~\cite{kuo2018_DiffuserCam_Fluorescence}. To fabricate, we repeatedly indent a copper block with a ball bearing of radius 7 mm. The indentations are made at random spacing (by hand) over an area larger than the 14.04 $\times$ 16.64 mm size of the PCO Edge 5.5 sensor. The result is a mold that is piecewise spherical with curvature matching the ball bearing. We use this block as a mold for UV-cured epoxy (Norland 61), with microscope slide on the top surface to ensure flatness. We then cure the epoxy and separate it from the mold. The epoxy has refractive index 1.56, yielding a diffuser with random lenslets of approximate focal length $f = 12.5 mm$. We mask the diffuser with a 13 $\times 15.5$ mm rectangular aperture, then mount the diffuser approximately 12.4 mm from the sensor. This results in magnification of $-.015\times$
 for objects placed $830$ mm away. 

\subsection{Artifacts due to time-varying FoV}\label{sec:sampling_and_aperture}
Given the structured sampling pattern of a rolling shutter sensor, we can reason about the system FoV geometrically. The set of scene points visible to each sensor pixel is determined by projecting rays from the pixel through the aperture. From this simple picture, we see that each pixel has a unique FoV. Because the rolling shutter pattern reads a band of rows simultaneously, this effectively means the FoV is varying with time: early in the exposure, the outer sensor rows are active, and cannot see the center of the FoV, while the inner rows (later frames) can. Because the sensor is blind to the on-axis points early in the exposure, these frames are determined via the regularizer. This explains the wiping artifact present in our videos in the early frames. If we were to use a single-shutter sensor, the effect would be more pronounced, as the FoV would sweep across the scene. This issue could be alleviated by distributing the active pixels more evenly across the sensor plane or by removing the aperture. In the current system, we simply discard the early frames of the video. In future builds, we could remove or enlarge the aperture, though this will preclude single-image calibration, and will lead to our shift-invariant lensless model breaking down at high angles. Such artifacts are correctable \cite{kuo2018_DiffuserCam_Fluorescence}, but lead to much slower processing times, and so we leave this for future work. 



\subsection{Limitations and future work}

For our prototype, there are two main limiting factors: the quality of the optics, and the CMOS sensor dynamics. Because the sensor's minimum exposure time limits the maximum usable frame rate, sensors with shorter exposure will perform better. Additionally, to match the line time to the exposure time, we would like to freely adjust the sensor's line timing; however, our sensor does not allow this. This leads us to use spatial downsampling as a workaround to effectively increase the line time to better match the band-limit. 

The second limiting factor is the quality of the diffuser. While our homemade diffusers are sufficient for proof-of-concept work, the resulting optics is fairly low quality, and the process is not well controlled. We can achieve the target focal length, but the focal spots (see Fig.~\ref{fig:caustics}) are extremely aberrated. This works well with the downsampling approach, as the caustics are not sharp enough to warrant using the full resolution sensor. However, to push our approach to the limit, we would need optics that can produce multiplexed PSFs with very sharp features. Coupled with a sensor capable of short exposures (on the order of the line time), our proposed architecture could achieve extremely high spatio-temporal resolution. For example, our current sensor can operate with line times as fast as $9.17$ $\mu s$, or over 100,000 fps.

Another limiting factor is the reduced measurement signal-to-noise caused by the multiplexing. Pushing this system to 100,000 fps would require exposure times shorter than 10 $\mu s$. Because the light from each point is distributed across the sensor with only a few pixels being recorded in each frame, this would require extremely bright scenes. Additionally, the combination of multiplexing and regularized reconstruction generally reduces the dynamic range of the recovered image, further limiting the method to high contrast scenes. 

As with most compressed sensing systems, it is difficult to validate the performance in general, since it is object dependent. We know from prior work~\cite{antipa2018diffusercam} that the performance degrades with scene complexity, and we observe this effect. While it does work for dense scenes, we require higher regularization, effectively limiting the usefulness for scenes that do not fit a gradient sparsity prior well. Introducing more sophisticated priors could mitigate this issue.

Our reconstructions are computationally expensive relative to a direct sampling approach. Achieving extremely short exposures and the fastest line time possible would require not downsampling the measurement, leading to a computationally expensive 3D inverse problem at gigavoxel scale. 

While we chose a dual-shutter camera for the experimental validation in this work, exploring the use of different programmable exposures could be extremely fruitful. Demonstrating the system with the more commonplace single shutter CMOS architecture would make it widely accessible, as the only other required equipment is a diffuser. Our current sensor also has a delay far longer than the line time between each sequential frame, preventing us from stringing together sequential frames into longer videos without a gap (see Supplementary Video 4~\cite{antipa_2019_ICCP_video_supplement}). A sensor that streamed continuously could alleviate this. It could also be useful to couple multiplexing optics with randomized sensor read patterns \cite{wei2018programmable_cmos}, as this will certainly lead to better video recovery.

\section*{Conclusion}
In conclusion, we have demonstrated that a spatially-multiplexing lensless camera can turn rolling shutter from a detriment into an advantage. We built a proof-of-concept system that resolves $1,500$ Hz dynamics at a frame rate of $4,545$ frames per second. We derived a theoretical temporal resolution bound based on our forward model, and confirmed our theoretical predictions with experiment. Our system relies on compressed sensing to solve an extremely underdetermined problem. We successfully observed samples with space-time bandwidth product far exceeding what could be observed with a direct sampling approach. Finally, we demonstrated our approach on a variety of fast-moving scenes, reliably recovering high speed videos from single rolling shutter images.

\section*{Acknowledgements}
The authors would like to thank The Moore Foundation, DARPA, and Bakar Fellows. This material is based upon work supported by the National Science Foundation under Grant No. 1617794. This work has also been supported by an Alfred P. Sloan Foundation fellowship. Emrah Bostan’s research is supported by the Swiss National Science Foundation (SNSF) under grant P2ELP2 172278. 

\bibliographystyle{IEEEtran}

\end{document}